\documentclass[
    ,final            
  ]
  {aipproc}

\usepackage{amsmath}

\layoutstyle{6x9}

\newcommand{\GeV}{\textrm{ GeV}}
\newcommand{\spa}{\ensuremath\text{SPS1a}^\prime}
\newcommand{\eq}[1]{eq.~\eqref{#1}}
\newcommand{\Order}[1]{\mathcal{O}\left(#1\right)}

\begin{document}

{\flushright \small 
 \vspace*{-1.5\baselineskip}
 Edinburgh 2006/33\\
 MPP-2006-137\\[-1.5\baselineskip]}

\title{Higgs-Boson Production in Association with Heavy Quarks}


\author{Wolfgang Hollik}{
  address={Max-Planck-Institut f\"ur Physik, M\"unchen, Germany}
}

\author{\underline{Michael Rauch}}{
  address={School of Physics, SUPA, University of Edinburgh, Scotland, UK}
}

\begin{abstract}
Associated production of a Higgs boson with a heavy, i.e.\ top or
bottom, quark--anti-quark pair provide observation channels for Higgs
bosons at the LHC which can be used to measure the respective Yukawa
couplings.  For the light supersymmetric Higgs boson we present SUSY-QCD
corrections at the one-loop level, which constitute a significant
contribution to the cross section.
\end{abstract}

\maketitle


\section{Introduction}

Higgs-boson Yukawa couplings to fermions are 
proportional to the fermion masses and hence are very small for the
light quarks, $u$, $d$, $s$ and $c$.
In contrast, the top-quark mass is of the same order as the Higgs vacuum
expectation value, leading to a top-quark Yukawa coupling 
close to $1$. The bottom-quark mass  also leads to a
rather weak Yukawa coupling in the Standard Model. In the
Minimal Supersymmetric Standard Model (MSSM), the coupling to the
lighter CP-even neutral Higgs boson $h^0$ can be enhanced for large values
of $\tan(\beta)$, the ratio of the two vacuum expectation values.
Such large Yukawa couplings make Higgs-boson production in association
with heavy quarks~\cite{Raitio:1978pt%
} 
a phenomenologically interesting process.
At $\Order{\alpha_s^2\alpha}$, the Higgs boson is 
emitted off one of the heavy-quark lines;  the cross
section is thus sensitive to the Yukawa coupling
and can be used to measure the respective Yukawa coupling. 
A precise  determination requires to include at least the next-order  
QCD corrections. Here we present   
for the case of MSSM Higgs bosons the results from a calculation
of the SUSY-QCD corrections,
supplementing the standard QCD corrections 
by the loop contributions with virtual gluinos and squarks.

\section{Bottom Quarks}

The production of a Higgs boson in association with a
bottom quark--anti-quark pair was intensively studied in the
literature~\cite{ATLAS-TDR,Dicus:1988cx%
}. The analysis
was soon extended~\cite{Carena:1998gk,Diaz-Cruz:1998qc%
}
to include the lightest MSSM-Higgs boson $h^0$. The diagram types are
exactly the same as in the Standard Model case;
only the bottom-quark--Higgs coupling is changed to its supersymmetric
counterpart. 
The standard QCD
corrections~\cite{Balazs:1998sb%
}
to this process are already known and reduce the dependence of the cross
section on the factorization and renormalization scales.  The
final-state bottom quarks are required to be explicitly observed in the
detector via $b$-tagging, in contrast to inclusive
processes~\cite{Harlander:2003ai} without $b$-tagging.
Therefore, a transverse-momentum cut on the
bottom-quark jets, typically $p_T \ge 20 \GeV$, is applied. 
The additional cuts
reduce the cross section by one or two orders of magnitude, but also
greatly reduce the background and ensure that the Higgs boson was
emitted from a bottom quark and is therefore proportional to the square
of the $b$-quark Yukawa coupling.

Here we concentrate on  the
SUSY-QCD corrections with squarks and gluinos in the loops.
Part of these corrections were already calculated in
ref.~\cite{Gao:2004wg}. There an effective $b\bar{b}h^0$-coupling
was used which includes the one-loop vertex corrections, but no box-type
or pentagon diagrams were added in their analysis. We have performed a
full one-loop calculation of the SUSY-QCD corrections.

In certain regions of the MSSM parameter space 
a large contribution to the SUSY-QCD corrections originates from the
effective coupling of the bottom quark to the second Higgs doublet.
This changes the relation between bottom-quark mass and Yukawa coupling
and the additional contribution is commonly referred to as $\Delta_b$
in the literature~\cite{Carena:1999py%
}. It is proportional to $\tan(\beta)$ and represents for large values of
$\tan(\beta)$ the dominant supersymmetric correction.
If the $\Delta_b$-contribution is compared with full one-loop results
it is necessary to include it only to one-loop order as well and not use
any resummed version, resulting in the replacement 
\begin{equation} 
  m_b \rightarrow m_b \left( 1-\Delta_b \right) \quad , 
\label{deltamb} 
\end{equation} 
which has been used when calculating $\Delta_b$-corrected tree-level
cross sections.

In order to assess the relative differences between cross sections
the following quantities have been defined.
The relative one-loop correction is given as
\begin{equation}
\Delta_1 = \frac{\sigma_1 - \sigma_{0}}{\sigma_0} \quad ,
\end{equation}
where $\sigma_0$ denotes the tree-level cross section and $\sigma_1$ the
one-loop one including SUSY-QCD corrections.
Additionally, a $\Delta_b$-corrected tree-level cross section
$\sigma_{\Delta}$ was
calculated by using the replacement of \eq{deltamb} and treating
the $\Delta_b$ term as a one-loop contribution. Additionally, the
contribution to the vertex from the term proportional to the second
mixing angle in the MSSM-Higgs sector, $\alpha$, was included in
$\sigma_{\Delta}$ according to
ref.~\cite{Carena:1998gk,Carena:1999py,Carena:1999bh%
}.
The relative correction using only these contributions is defined as
\begin{equation}
\Delta_{\tilde{0}} =
  \frac{\sigma_{\Delta} - \sigma_{0}}{\sigma_0} \quad .
\end{equation}

The Feynman diagrams were generated using
FeynArts~\cite{Kublbeck:1990xc}, the matrix elements
calculated by FormCalc~\cite{Hahn:1998yk} and
the loop integrals numerically evaluated by
LoopTools~\cite{Hahn:2006qw}. The
convolution with the parton distribution functions was performed with
HadCalc~\cite{Rauch:HadCalc} using the PDF set of
ref.~\cite{Martin:2002aw}.

\begin{table}
\begin{tabular}{l|r|r|r|r||l|r|r|r}
Process & $\sigma_0$ [fb]
& $\sigma_1$ [fb]
&  $ \Delta_1 $ [\%] & $ \Delta_{\tilde{0}} $ [\%] &
Process & $\sigma_0$ [fb]
& $\sigma_1$ [fb]
&  $ \Delta_1 $ [\%] \\\hline
$d\bar{d}\rightarrow b\bar{b}h^0$ & $0.107$ & $0.104$ & $-2.48$ & $-1.95$ &
$d\bar{d}\rightarrow t\bar{t}h^0$ & $42.7$ & $37.6$ & $-11.77$ \\
$u\bar{u}\rightarrow b\bar{b}h^0$ & $0.168$ & $0.164$ & $-2.56$ & $-1.95$ &
$u\bar{u}\rightarrow t\bar{t}h^0$ & $71.9$ & $63.4$ & $-11.81$ \\
$s\bar{s}\rightarrow b\bar{b}h^0$ & $0.028$ & $0.028$ & $-2.26$ & $-1.95$ &
$s\bar{s}\rightarrow t\bar{t}h^0$ & $7.5$ & $6.6$ & $-11.58$ \\
$c\bar{c}\rightarrow b\bar{b}h^0$ & $0.013$ & $0.012$ & $-2.20$ & $-1.95$ &
$c\bar{c}\rightarrow t\bar{t}h^0$ & $2.8$ & $2.5$ & $-11.53$ \\
$gg\rightarrow b\bar{b}h^0$ & $35.647$ & $33.734$ & $-5.37$ & $-1.95$ &
$gg\rightarrow t\bar{t}h^0$ & $273.7$ & $264.7$ & $-3.30$ \\\hline
$pp\rightarrow b\bar{b}h^0$ & $35.963$ & $34.042$ & $-5.34$ & $-1.95$ &
$pp\rightarrow t\bar{t}h^0$ & $399.0$ & $374.8$ & $-5.96$
\end{tabular}
\caption{Hadronic cross sections for $b\bar{b}h^0$ and 
  $t\bar{t}h^0$ production at the parameter point $\spa$.}
\label{hq:bspa}
\end{table}

The left-hand side of Table~\ref{hq:bspa} contains the individual
contributions from the various partonic processes
to the hadronic process
$pp \rightarrow b\bar{b}h^0$ 
and their sum, for the
MSSM reference point $\spa$~\cite{spa}. The gluon-fusion process
clearly dominates the total hadronic cross section. This is because for
the quark--anti-quark annihilation diagrams only an $s$-channel topology
exists, which is propagator-suppressed. For the gluon-fusion diagrams
there is an additional $t$-channel diagram which does not suffer from such
a suppression. Additionally
one can see that the $\Delta_b$-corrected tree-level cross section
accounts only for less than half of the total SUSY-QCD corrections for
this parameter point, and therefore a full calculation is necessary to
determine the size of the additional contribution. The details of this 
additional contribution will be discussed in a future 
publication~\cite{WHMR:future}.

\section{Top Quarks}

The production of a Higgs boson in association with a top
quark--anti-quark
pair~\cite{Kunszt:1984ri%
}
proceeds in the same way as the one with a bottom quark--anti-quark pair
and the same Feynman diagrams appear, where the bottom-quark line is
replaced by a top-quark line. The standard QCD corrections to
this process are also available in the
literature~\cite{Beenakker:2001rj%
}.
A calculation of the SUSY-QCD corrections was performed recently
in ref.~\cite{Peng:2005ti}. As the figures of this article include both
standard and SUSY-QCD contributions a direct
comparison of the numerical results is difficult. 
The principal behavior when varying MSSM parameters agrees.
Our calculation was performed using the same tools as mentioned
beforehand in the bottom-quark case. No cuts were applied to the final
state. 

On the right-hand side of Table~\ref{hq:bspa} the results for the MSSM
parameter point $\spa$ are presented. In this case the gluon-fusion
contribution is still the
dominant one, but also the quark--anti-quark--annihilation subprocesses
reach a significant size and cannot be neglected any more. This is
because to produce the final state a higher center-of-mass energy than
for bottom quarks is needed. The rapid decrease of the gluon density in
the proton with growing momentum fraction $x$ partly cancels the effect
of the s-channel propagator suppression in quark--anti-quark
annihilation. We find that the total size of the SUSY-QCD corrections is
of the order of several percent.

\section{Summary}
Higgs-boson production in association with heavy, i.e.\ bottom or top,
quarks is an important way to measure the respective Yukawa couplings.
In the MSSM besides the standard QCD corrections also SUSY-QCD
corrections appear. They modify the total cross section significantly
and should be taken into account to extract the Yukawa coupling precisely 
from future experimental data.


\begin{theacknowledgments}
  We would like to thank T Plehn for careful reading of the manuscript.
  The work of MR was supported by the Scottish Universities Physics
  Alliance (SUPA).
\end{theacknowledgments}





\begin{thebibliography}{43}
\expandafter\ifx\csname natexlab\endcsname\relax\def\natexlab#1{#1}\fi
\providecommand{\enquote}[1]{``#1''}
\expandafter\ifx\csname url\endcsname\relax
  \def\url#1{\texttt{#1}}\fi
\expandafter\ifx\csname urlprefix\endcsname\relax\def\urlprefix{URL }\fi
\providecommand{\eprint}[2][]{\url{#2}}

\bibitem[Raitio and Wada(1979)]{Raitio:1978pt}
R.~Raitio, and W.~W. Wada, \emph{Phys. Rev.} \textbf{D19}, 941 (1979);
%
A.~S. Bagdasaryan, R.~S. Egorian, S.~G. Grigorian, and S.~G. Matinyan,
  \emph{Sov. J. Nucl. Phys.} \textbf{46}, 315 (1987);
%
J.~N. Ng, and P.~Zakarauskas, \emph{Phys. Rev.} \textbf{D29}, 876 (1984).

\bibitem[{ATLAS Collaboration}(1999)]{ATLAS-TDR}
{ATLAS Collaboration}  (1999), \eprint{CERN/LHCC 99-14};
J.~Cammin, and M.~Schumacher, \eprint{ATL-PHYS-2003-024};
S.~Cucciarelli {\it et al.}, \eprint{CERN-CMS-NOTE-2006-119}.

\bibitem[Dicus and Willenbrock(1989)]{Dicus:1988cx}
D.~A. Dicus, and S.~Willenbrock, \emph{Phys. Rev.} \textbf{D39}, 751 (1989);
%
J.~Campbell, et~al.  (2004), \eprint{hep-ph/0405302}.

\bibitem[Carena et~al.(1999)]{Carena:1998gk}
M.~Carena, S.~Mrenna, and C.~E.~M. Wagner, \emph{Phys. Rev.} \textbf{D60},
  075010 (1999), \eprint{hep-ph/9808312}.

\bibitem[Diaz-Cruz et~al.(1998)]{Diaz-Cruz:1998qc}
Z.~Kunszt, and F.~Zwirner, \emph{Nucl. Phys.} \textbf{B385}, 3--75 (1992),
  \eprint{hep-ph/9203223};
%
J.~Dai, J.~F. Gunion, and R.~Vega, \emph{Phys. Lett.} \textbf{B345}, 29--35
  (1995), \eprint{hep-ph/9403362},
%
\emph{Phys. Lett.} \textbf{B387}, 801--803
  (1996), \eprint{hep-ph/9607379};
%
E.~Richter-Was, and D.~Froidevaux, \emph{Z. Phys.} \textbf{C76}, 665--676
  (1997), \eprint{hep-ph/9708455};
J.~L. Diaz-Cruz, H.-J. He, T.~Tait, and C.~P. Yuan, \emph{Phys. Rev. Lett.}
  \textbf{80}, 4641--4644 (1998), \eprint{hep-ph/9802294};
%
C.~Balazs, J.~L. Diaz-Cruz, H.~J. He, T.~Tait, and C.~P. Yuan, \emph{Phys.
  Rev.} \textbf{D59}, 055016 (1999{\natexlab{a}}), \eprint{hep-ph/9807349}.
%

\bibitem[Balazs et~al.(1999{\natexlab{b}})]{Balazs:1998sb}
C.~Balazs, H.-J. He, and C.~P. Yuan, \emph{Phys. Rev.} \textbf{D60}, 114001
  (1999{\natexlab{b}}), \eprint{hep-ph/9812263};
%
D.~Dicus, T.~Stelzer, Z.~Sullivan, and S.~Willenbrock, \emph{Phys. Rev.}
  \textbf{D59}, 094016 (1999), \eprint{hep-ph/9811492};
%
S.~Dittmaier, M.~Kramer, and M.~Spira, \emph{Phys. Rev.} \textbf{D70}, 074010
  (2004), \eprint{hep-ph/0309204};
%
S.~Dawson, C.~B. Jackson, L.~Reina, and D.~Wackeroth, \emph{Phys. Rev.}
  \textbf{D69}, 074027 (2004), \eprint{hep-ph/0311067},
%
  (2005),
  \eprint{hep-ph/0508293}.

\bibitem[Harlander et~al.(2003)]{Harlander:2003ai}
R.~V.~Harlander and W.~B.~Kilgore, \emph{Phys. Rev.} \textbf{D68},
  013001 (2003), \eprint{hep-ph/0304035}.

\bibitem[Gao et~al.(2005)]{Gao:2004wg}
G.~Gao, R.~J. Oakes, and J.~M. Yang, \emph{Phys. Rev.} \textbf{D71}, 095005
  (2005), \eprint{hep-ph/0412356}.

\bibitem[Carena et~al.(2000{\natexlab{a}})]{Carena:1999py}
M.~Carena, M.~Olechowski, S.~Pokorski and C.~E.~M.~Wagner, \emph{Nucl. Phys.}
  \textbf{B426}, 269--300 (1994), \eprint{hep-ph/9402253};
L.~J.~Hall, R.~Rattazzi and U.~Sarid, \emph{Phys. Rev.} 
  \textbf{D50}, 7048--7065 (1994), \eprint{hep-ph/9306309};
M.~Carena, D.~Garcia, U.~Nierste, and C.~E.~M. Wagner, \emph{Nucl. Phys.}
  \textbf{B577}, 88--120 (2000{\natexlab{a}}), \eprint{hep-ph/9912516};
%
J.~Guasch, W.~Hollik, and S.~Penaranda, \emph{Phys. Lett.} \textbf{B515},
  367--374 (2001), \eprint{hep-ph/0106027};
%
J.~Guasch, P.~Hafliger, and M.~Spira, \emph{Phys. Rev.} \textbf{D68}, 115001
  (2003), \eprint{hep-ph/0305101}.

\bibitem[Carena et~al.(2000{\natexlab{b}})]{Carena:1999bh}
M.~Carena, S.~Mrenna, and C.~E.~M. Wagner, \emph{Phys. Rev.} \textbf{D62},
  055008 (2000{\natexlab{b}}), \eprint{hep-ph/9907422}.

\bibitem[Kublbeck et~al.(1990)]{Kublbeck:1990xc}
J.~Kublbeck, M.~Bohm, and A.~Denner, \emph{Comput. Phys. Commun.} \textbf{60},
  165--180 (1990).

\bibitem[Hahn and Perez-Victoria(1999)]{Hahn:1998yk}
T.~Hahn, and M.~Perez-Victoria, \emph{Comput. Phys. Commun.} \textbf{118},
  153--165 (1999), \eprint{hep-ph/9807565}.

\bibitem[Hahn and Rauch(2006)]{Hahn:2006qw}
T.~Hahn, and M.~Rauch, \emph{Nucl. Phys. Proc. Suppl.} \textbf{157}, 236--240
  (2006), \eprint{hep-ph/0601248}.

\bibitem[Rauch(2006)]{Rauch:HadCalc}
M.~Rauch, \emph{PhD thesis}, (2006),
  \eprint{urn:nbn:de:bvb:91-diss20060502-1522300621}.

\bibitem[Martin et~al.(2003)]{Martin:2002aw}
A.~D. Martin, R.~G. Roberts, W.~J. Stirling, and R.~S. Thorne, \emph{Eur. Phys.
  J.} \textbf{C28}, 455--473 (2003), \eprint{hep-ph/0211080}.

\bibitem[Aguilar-Saavedra et~al.(2005)]{spa}
J.~A. Aguilar-Saavedra, et~al.  (2005), \eprint{hep-ph/0511344}.

\bibitem{WHMR:future}
W.~Hollik, M.~Rauch, (in preparation).

\bibitem[Kunszt(1984)]{Kunszt:1984ri}
Z.~Kunszt, \emph{Nucl. Phys.} \textbf{B247}, 339 (1984);
%
W.~J. Marciano, and F.~E. Paige, \emph{Phys. Rev. Lett.} \textbf{66},
  2433--2435 (1991);
%
J.~F. Gunion, \emph{Phys. Lett.} \textbf{B261}, 510--517 (1991);
%
J.~Goldstein, et~al., \emph{Phys. Rev. Lett.} \textbf{86}, 1694--1697 (2001),
  \eprint{hep-ph/0006311};

\bibitem[Beenakker et~al.(2001)]{Beenakker:2001rj}
W.~Beenakker, et~al., \emph{Phys. Rev. Lett.} \textbf{87}, 201805 (2001),
  \eprint{hep-ph/0107081},
%
\emph{Nucl. Phys.} \textbf{B653}, 151--203 (2003),
  \eprint{hep-ph/0211352};
%
L.~Reina, and S.~Dawson, \emph{Phys. Rev. Lett.} \textbf{87}, 201804 (2001),
  \eprint{hep-ph/0107101};
%
S.~Dawson, L.~H. Orr, L.~Reina, and D.~Wackeroth, \emph{Phys. Rev.}
  \textbf{D67}, 071503 (2003), \eprint{hep-ph/0211438}.

\bibitem[Peng et~al.(2005)]{Peng:2005ti}
W.~Peng, et~al.  (2005), \eprint{hep-ph/0505086}.

\end{thebibliography}

\end{document}